\documentclass[12pt]{article}
\usepackage{amsmath,bm,graphicx,epsfig}
\usepackage{hyperref}
\include{epsf} 
\newcommand\as{\alpha_S} 
\def\ltap{\raisebox{-.6ex}{\rlap{$\,\sim\,$}} \raisebox{.4ex}{$\,<\,$}} 
\def\gtap{\raisebox{-.6ex}{\rlap{$\,\sim\,$}} \raisebox{.4ex}{$\,>\,$}} 

\def\nn{\nonumber} 
 
\def\beqn{\begin{eqnarray}} 
\def\eeqn{\end{eqnarray}} 
\def\beq{\begin{equation}} 
\def\eeq{\end{equation}}

\def\pla{\lambda'} 
\newcommand\f[2]{\frac{#1}{#2}} 

\begin{document}
\begin{titlepage}
\begin{flushright}
TIF-UNIMI-2018-4\\
\end{flushright}

\renewcommand{\thefootnote}{\fnsymbol{footnote}}
\vspace*{2.cm}

\begin{center}
{\Large \bf 
  Combining QED and QCD\\[0.1cm] transverse-momentum resummation\\[0.1cm]
  for \boldmath$Z$ boson  production at hadron colliders\\[0.1cm] 
%Transverse-momentum resummation \\
%for Higgs boson pair production at the LHC  \\ with top-quark mass effects\\ 
}
\end{center}

\par \vspace{6mm}
\begin{center}
  {\bf Leandro Cieri${}^{(a)}$, Giancarlo Ferrera${}^{(b)}$} and
  {\bf Germ\'an F.\,R.\ Sborlini${}^{(b)}$}\\

\vspace{8mm}

${}^{(a)}$ 
{\itshape  INFN, Sezione di Milano-Bicocca,\\
Piazza della Scienza 3, I-20126 Milan, Italy}\\\vspace{3mm}

${}^{(b)}$ 
{\itshape Tif Lab,
Dipartimento di Fisica, Universit\`a di Milano and\\ INFN, Sezione di Milano,
Via Celoria 16, I-20133 Milan, Italy}\\\vspace{1mm}
\end{center}

\vspace{1.25cm}

%\pacs{13.87.Ce,12.38Bx}

\par \vspace{2mm}
\begin{center} {\large \bf Abstract} \end{center}
\begin{quote}
\pretolerance 10000
We consider the transverse-momentum ($q_T$) distribution of
$Z$ bosons produced in hadronic collisions.
At small values of $q_T$, we perform the analytic resummation
of the logarithmically enhanced QED contributions %,
%due to soft and/or collinear photon emissions,
up to next-to-leading logarithmic accuracy, including 
the mixed QCD-QED contributions at leading logarithmic accuracy. %,
Resummed results are consistently matched with the next-to-leading
fixed-order results (i.e.\ $\mathcal{O}(\alpha^2)$)
at small, intermediate and large values of $q_T$.
We combine the QED corrections with the known QCD results at
next-to-next-to-leading order ($\mathcal{O}(\alpha_S^2)$) and
next-to-next-to-leading logarithmic accuracy. 
We show numerical results at LHC and Tevatron energies,
studying the impact of the QED corrections and providing
an estimate of the corresponding perturbative uncertainty.
Our analytic results for the combined QED and QCD resummation,
obtained through an extension of the $q_T$ resummation formalism in QCD,
are valid for
the production of generic neutral and colourless high-mass systems
in hadronic collision.

\end{quote}

\vspace*{\fill}
\vspace*{2.2cm}

\begin{flushleft}
May 2018
\end{flushleft}
\end{titlepage}

\setcounter{footnote}{1}
\renewcommand{\thefootnote}{\fnsymbol{footnote}}

\section{Introduction}

The production of vector bosons ($W$ and $Z/\gamma^*$), through the
Drell-Yan (DY) mechanism~\cite{Drell:1970wh}, is one of the most prominent
hard-scattering processes %for physics studies
at high-energy hadron colliders,
such as the Tevatron and the LHC.
In particular, it provides strong tests of the Standard Model (SM) and of QCD,
it is important for precise measurements of the mass and the width of the $W$ boson and
for the extraction of the electroweak (EW) mixing angle, it gives stringent
information on the parton densities functions of the colliding hadrons, and
it represents an important background to other processes and to various
possible signals from physics beyond the SM.

Owing to the above reasons, it is of primary importance to provide accurate
theoretical predictions for vector-boson production cross sections
and related kinematical distributions. This task requires
detailed computations of the higher-order radiative corrections in 
QCD~\cite{Hamberg:1990np}%,Anastasiou:2003ds,Melnikov:2006di,Catani:2009sm,Catani:2010en,Boughezal:2015dva,
--\cite{Ridder:2015dxa} and
in the EW theory~\cite{ewW}--\cite{Dittmaier:2015rxo}.

Among the various kinematical distributions, the transverse-momentum ($q_T$)
spectrum of the vector boson has a particularly high relevance.
A good understanding of the $Z$ boson $q_T$ distribution gives important
information on the $W$ boson production mechanism. In turn,
an accurate theoretical prediction of the $q_T$ distribution of the $W$
boson, in the intermediate and small $q_T$ region, is crucial for a precise
measurement of the $W$ mass~\cite{CarloniCalame:2016ouw,Aaltonen:2013iut,Aaboud:2017svj}.
In order to fully exploit the high precision of the LHC measurements,
corresponding theoretical predictions  with increasing accuracy are therefore essential.

In the large-$q_T$ region, where the transverse momentum is of the order of the vector boson mass $M$,
fixed-order perturbative calculations are theoretically 
justified~\cite{Ellis:1981hk,Arnold:1988dp,Gonsalves:1989ar,Gehrmann-DeRidder:2017mvr}. However
the bulk of the vector boson cross section lies in the small-$q_T$ region ($q_T\ll M$)
where the reliability of the fixed-order expansion is spoiled by the presence of
large logarithmic corrections of the type $\ln (M^2/q_T^2)$
due to the initial state radiation of soft and/or collinear partons.

In order to obtain reliable predictions, 
the enhanced-logarithmic terms have to be evaluated and systematically
resummed to all orders in perturbation theory. The formalism to resum these large logarithmic
corrections in QCD has been developed from the late seventies~\cite{Dokshitzer:hw}--\cite{Catani:2013tia}
starting from results settled for QED~\cite{Sudakov:1954sw}.

While the resummation of the large logarithmic corrections in QCD in the small $q_T$ region is mandatory,
the impact of the large logarithmic QED corrections from the initial states
is expected to be substantial only at extremely small values of $q_T$
due to the  relative smallness of the electromagnetic coupling 
$\alpha$ with respect to the strong coupling $\alpha_S$.
 %of the type $\alpha^n \ln (M^2/q_T^2)$ 
%for the $q_T$ distribution 

Nonetheless all-order resummation of QED  emissions  can have non-negligible impact on pure QCD resummed results.
Moreover
%Although the QED effects are suppressed %with respect to the QCD ones 
%by the relative smallness of the electromagnetic coupling $\alpha$ with respect to the strong coupling $\alpha_S$, 
%we observe that 
%since 
%within fixed-order 
%perturbation theory 
%being
since the first-order QCD contribution at finite value of $q_T$ is an
$\mathcal{O}(\alpha_S)$ correction relative to the Born-level result,
the QED contribution corresponds to  
an $\mathcal{O}(\alpha/\alpha_S)$ correction to the QCD result. 
Given the present accuracy of the theoretical predictions~\cite{Alioli:2016fum} and the high-precision experimental LHC data, 
it is 
therefore 
important to quantify in a consistent way the impact of QED contributions for the  $q_T$ distribution of vector bosons.

In this paper we extend the QCD  formalism of Refs.~\cite{Catani:2000vq,Bozzi:2005wk}
in order to combine QCD and QED transverse momentum resummation.
In particular we explicitly consider the case of $Z$ production at the Tevatron and the LHC by performing %the %combined
QED and QCD resummation %of QED and QCD emissions 
up to next-to-leading logarithmic (NLL) and
at next-to-next-to-leading logarithmic (NNLL) accuracy, respectively. We also include the resummation of the
mixed QCD-QED corrections at leading logarithmic accuracy.
In order to achieve a uniform theoretical accuracy for the entire range of transverse momenta,
our resummed calculation  has been consistently 
matched at small and intermediate-large values of $q_T$
with the fixed-order corrections at next-to-leading order (NLO) in QED (i.e.\ $\mathcal{O}(\alpha^2)$)
and next-to-next-to-leading order (NNLO) in QCD (i.e.\ $\mathcal{O}(\alpha_S^2)$).
We show that the QED corrections have a non negligible impact on the 
$q_T$ spectrum of the $Z$ boson.
%(with respect to the corresponding QCD results) 
In particular the inclusion of the NLL+NLO QED corrections reduces the effect 
of renormalization scale ambiguity of
the QED coupling $\alpha$  by consistently including
 running coupling  effects.

The paper is organized as follows. In Sect.~\ref{sec:computation} we present the combined QED and QCD resummation formalism,
which is valid for a generic process involving a
colorless and neutral final state (namely, $Z/\gamma^*$, $ZZ$, $\gamma\gamma$, Higgs boson(s), etc.),
providing the explicit expression for the universal coefficients necessary to include
QED resummation up to NLL accuracy.
%Altough we consider the specific case of the specific case of $Z$ bosons production,
%the formalism we present is valid for a generic process involving a
%colorless and neutral final state (namely, $Z/gamma^*$, $ZZ$, $\gamma\gamma$, Higgs boson(s), etc.).
%Our notation closely follows the one introduced in the orginal references for soft gluon
%emission~\cite{Catani:2000vq,Bozzi:2005wk,Bozzi:2007pn}.
% and for the $Z/\gamma^*$ interference.
In Sect.~\ref{sec:results} we show illustrative numerical results for $Z$ boson production at Tevatron and LHC energies. In particular,
we present combined resummation results up to NLL+NLO in QED and NNLL+NNLO in QCD, quantifying the impact of QED effects 
with respect to the corresponding NNLL+NNLO QCD
results. We then perform a study of the scale dependence of the 
QED corrections in order to estimate the
corresponding perturbative uncertainty.
In Sect.~\ref{sec:conclusions} we summarize our results.
%The Appendix ~\ref{app} presents with detail the QED $q_T$ resummation coefficients that are necessary for a calculation up to NLO+NLL.

\section{Combined QED and QCD transverse-momentum resummation}
\label{sec:computation}
In this Section we extend the QCD transverse-momentum resummation  formalism of Refs.~\cite{Catani:2000vq,Bozzi:2005wk} in order to consistently include
the resummation of QED perturbative logarithmic corrections
in the generic case of the production of 
high-mass systems composed by colourless and neutral particles
in hadronic collision.
In particular we highlight the differences that arise when considering the inclusion of the QED emissions, by considering the specific case
of the hadroproduction of an on-shell $Z$ boson: $h_1 + h_2 \to F +X$, 
($F=Z$). 
%A detailed discussion of the $q_T$ resummation formalism in QCD used 
%in this paper can be found in Refs.~\cite{}.

%~\footnote{} 

%In order to implement the QED effects in the neutral vector boson production, we rely on the original \texttt{DYqt} \cite{Bozzi:2008bb,Bozzi:2010xn} code and properly introduce the modifications for the associated QED contributions. Within this code, the total cross-section is separated into two pieces, the resummed and the fixed order components, respectively. Then, by performing a proper power-expansion, both corrections are merged into the so-called matched prediction, which includes the low $q_T$ corrections due to soft gluon emissions, as well as the perturbative terms that dominate the higher-order behavior in the high $q_T$ region. 

We start our presentation of combined QCD and QED transverse-momentum
resummation by briefly recalling the main features of $q_T$ resummation
for QCD~\cite{Catani:2000vq,Bozzi:2005wk}.
According to the QCD factorization theorem, the transverse-momentum 
differential cross section for this process 
can be written as
%~\footnote{If %$F=\gamma^*$ or if 
%$F$ is produced `off-shell', 
%$d\sigma/{d q_T^2}$ actually denotes the differential cross section $M^2d\sigma/{dM^2d q_T^2}$, where $M$ is the invariant mass of $F$.}
\beq
\label{eq1}
\frac{d \sigma_{h_1 h_2 \to F}}{d q_T^2}(q_T,M,s) = \sum_{a_1,a_2}\int_0^1 dx_1 \int_0^1 dx_2\,  f_{a_1/h_1}(x_1,\mu_F^2) f_{a_2/h_2}(x_2,\mu_F^2)
\frac{d \hat{\sigma}_{a_1 a_2 \to F}}{d q_T^2}(q_T,M,\hat s;%\mu_R^2,
\mu_F^2) \, ,
\eeq
where $f_{a/h}(x,\mu_F^2)$ ($a=q,\bar q, g$) are the parton densities of the hadron $h$ at the factorization scale $\mu_F$, $d{\hat\sigma_{a_1 a_2 \to F}}$ are 
the partonic cross sections and $s$ ($\hat s=x_1x_2 s$) is the hadronic (partonic) centre--of--mass energy. %, and $\mu_R$ is the factorization scale. 

The partonic cross sections $d{\hat\sigma/dq_T^2}$ in Eq.\,(\ref{eq1}) 
can be decomposed as follows: 
$d \hat{\sigma}=d \hat{\sigma}^{({\rm res.})}+ d \hat{\sigma}^{({\rm fin.})} \,$, 
where the `resummed' component ($d \hat{\sigma}^{({\rm res.})}$)
contains all the logarithmically enhanced contributions at small $q_T$ 
proportional to $\delta(q_T^2)$ or to $1/q_T^2\ln^m(M^2/q_T^2)$ 
and the `finite' component ($d \hat{\sigma}^{({\rm fin.})}$) is free of such 
contributions.

The resummation procedure is carried out in the impact-parameter ($b$) space conjugated to $q_T$. The resummed component is then obtained
by performing the inverse Bessel transformation with respect to $b$:
\beq
\frac{d \hat{\sigma}_{a_1 a_2 \to F}^{({\rm res.})}}{d q_T^2} (q_T,M,\hat s;\mu_F)= 
\frac{M^2}{\hat{s}} \int_0^{\infty} \, db \frac{b}{2} J_0(b \, q_T) \, 
{\cal W}^F_{a_1a_2}(b,M,\hat{s};\mu_F) \, ,
\eeq
where $J_0(x)$ is the 0th-order Bessel function. The resummation structure of  ${\cal W}^F_{a_1a_2}$ can be organized in an exponential form by considering the 
Mellin $N$-moments ${\cal W}^F_{a_1a_2,N}$~\footnote{The $N$-moments of
a generic function $h(z)$ are defined as $h_N=\int_0^1 dz\, z^{N-1}h(z)$.} with respect to $z=M^2/\hat s$ at fixed $M$. In the simplified flavour-diagonal case
($a_1a_2=c\bar c$) it reads: 
%\beq
%{\cal W}^F_{a_1a_2,N}(b,M,\hat{s};\mu_F^2) = \int_0^1 \, dz \, z^{N-1} \, {\cal W}^F_{a_1b_2}(b,M,\hat{s}=M^2/z;\mu_F^2) \, . 
%\eeq 
\beq
\label{eqW}
      {\cal W}^F_{N}(b,M;\mu_F) =
      \hat\sigma_{F}^{(0)}(M)\,{\cal H}_{N}^F(\as;M^2/\mu_R^2,M^2/\mu_F^2,M^2/Q^2)
\times  \exp \left\{ {\cal G}_{N}(\as, L; M^2/\mu_R^2,M^2/Q^2) \right\} \,,
\eeq
where $\hat\sigma_{F}^{(0)}$ is the lowest-order partonic cross section, $L=\log(Q^2b^2/b_0^2+1)$ 
is the logarithmic expansion parameter, $Q$ is the resummation scale~\cite{Bozzi:2005wk} and $b_0=2 e^{-\gamma_E}$ 
(with $\gamma_E=0.5772\dots$ the Euler-Mascheroni constant). In Eq.\,(\ref{eqW}) we explicitly introduced the renormalization scale $\mu_R$
and the perturbative QCD coupling $\as=\as(\mu_R^2)$.

The universal %{\itshape universal} (process independent) 
form factor $\exp\left\{{\cal G}_N\right\}$ includes (and resums to all orders) the 
large logarithmic terms $\as^nL^m$
($1\leq m \leq 2n $). %which diverges order-by-order in $\as$ as $b\to \infty$
It can be systematically expanded in powers of $\as$ as follows:
\begin{align}
\label{eqG}
{\cal G}_{N}(\as, L%;M^2/\mu^2_R,M^2/Q^2
)&=L \;g^{(1)}(\as L)+g_N^{(2)}(\as L) %;M^2/\mu_R^2,M^2/Q^2)\nn\\
+\frac{\as}{\pi} g_N^{(3)}(\as L) %;M^2/\mu_R^2,M^2/Q^2)\nn\\
+\sum_{n=4}^\infty \left(\frac{\as}{\pi}\right)^{n-2} g_N^{(n)}(\as L)\,, %;M^2/\mu_R^2,M^2/Q^2)\;,
\end{align}
where the term $L\, g^{(1)}$ collects the leading logarithmic (LL) 
contributions, the function $g_N^{(2)}$ controls the NLL contributions, $g_N^{(3)}$ includes the NNLL terms and so forth.

All the perturbative terms that behave as constants in the limit $b\to\infty$ 
are included in the process dependent  hard-collinear function ${\cal H}_N^{F}$ which
has a customary %fixed-order 
perturbative expansion: 
\begin{align}
\label{eqH}
{\cal H}_N^{F}(\as)= %;M^2/\mu^2_R,M^2/\mu^2_F,M^2/Q^2)&=
1+ \frac{\as}{\pi} \,{\cal H}_N^{F \,(1)} %(M^2/\mu^2_R,M^2/\mu^2_F,M^2/Q^2) \nn \\
+ 
\left(\frac{\as}{\pi}\right)^2 
\,{\cal H}_N^{F \,(2)} % (M^2/\mu^2_R,M^2/\mu^2_F,M^2/Q^2)\nn\\
+\sum_{n=3}^\infty \left(\frac{\as}{\pi}\right)^{n}\,{\cal H}_N^{F \,(n)} \;\;. % (M^2/\mu^2_R,M^2/\mu^2_F,M^2/Q^2) \;\;.
\end{align}

We now discuss how to extend the QCD resummation formalism in order to include and consistently resum the large logarithmic QED corrections
of the type $\alpha^n L^m$ ($1\leq m \leq 2n $). 
We start form Eq.\,(\ref{eq1}) and we consider also the inclusion of the photon
parton density $f_{\gamma/h}(x,\mu_F^2)$.
The  Eq.\,(\ref{eqW}) have then  to be  replaced by the following generalized expressions~\footnote{In order to simplify the expressions
  in this Section we do not explicitly introduce an additional dependence on perturbative scales in QED
  (i.e.\ we fix them to be equal to the corresponding QCD ones).}:
\beq
\label{eqW2}
{\cal W}_{N}'^F(b,M;\mu_F) = \hat\sigma_{F}^{(0)}(M)\,{\cal H}_{N}'^F(\as,\alpha;M^2/\mu_R^2,M^2/\mu_F^2,M^2/Q^2)
\times  \exp \left\{ {\cal G}_{N}'(\as,\alpha, L; M^2/\mu_R^2,M^2/Q^2) \right\} \,,
\eeq
where the form factor ${\cal G}_{N}'(\as,\alpha, L)$ in Eq.\,(\ref{eqW2}) has a double perturbative expansion
in powers of $\as$ and $\alpha$
\begin{align}
\label{eqG2}
{\cal G}_{N}'(\as,\alpha, L)&={\cal G}_{N}(\as, L)
+
L \; g'^{(1)}(\alpha L)+ g_N'^{(2)}(\alpha L) 
+\sum_{n=3}^\infty \left(\frac{\alpha}{\pi}\right)^{n-2} g_N'^{(n)}(\alpha L)\nn\\
&
+
g'^{(1,1)}(\as L,\alpha L)
+
\sum_{n,m=1 \atop n+m\neq 2}^\infty \left(\frac{\as}{\pi}\right)^{n-1}\left(\frac{\alpha}{\pi}\right)^{m-1} g_N'^{(n,m)}(\as L,\alpha L)\;\;,
%{\cal G}_{N}(\as, L)+
%L \;g^{(1)}(\as L)+g_N^{(2)}(\as L)\nn\\
%&+\frac{\as}{\pi} g_N^{(3)}(\as L)+\dots\;,
\end{align}
where $\alpha=\alpha(\mu_R)$ is the electromagnetic coupling evaluated at the renormalization scale 
$\mu_R$.
%~\footnote{In order to simplify the expressions in this Section
%we consider $\tilde\mu_R=\mu_R$.}.
The term $L\, g'^{(1)}$ collects the LL 
contributions in QED, the function $g_N'^{(2)}$ controls the NLL QED and so forth, while the terms
$g'^{(1,1)}(\as L,\alpha L)$  and $g_N'^{(n,m)}(\as L,\alpha L)$  include respectively the leading and subleading mixed QCD-QED corrections.
The extension of the Eq.\,(\ref{eqH}) has an analogous double perturbative expansion in powers of $\as$ and $\alpha$:
\begin{align}
\label{eqH2}
{\cal H}_N'^{F}(\as,\alpha)&= {\cal H}_N^{F}(\as)
+ \frac{\alpha}{\pi} \,{\cal H}_N'^{F \,(1)}
+\sum_{n=2}^\infty \left(\frac{\alpha}{\pi}\right)^{n}\,{\cal H}_N'^{F \,(n)} \nn\\ 
&+\sum_{n,m=1}^\infty \left(\frac{\alpha_S}{\pi}\right)^{n}\left(\frac{\alpha}{\pi}\right)^{m}\,{\cal H}_N'^{F \,(n,m)}\,,
% (M^2/\mu^2_R,M^2/\mu^2_F,M^2/Q^2) \;\;.
\end{align}
where the pure QED corrections are controlled by 
the coefficients ${\cal H}_N'^{F \,(n)}$ while the mixed QCD-QED ones
are contained in the coefficients ${\cal H}_N'^{F \,(n,m)}$.

%The functions $g'^{(n)}_N$ and ${\cal H}_N'^{F \,(n)}$ can be obtained from the corresponding ones in QCD,  through an Abelianization 
%algorithm~\cite{deFlorian:2015ujt,deFlorian:2016gvk}.
The LL and NLL functions $g'^{(1)}$ and $g_N'^{(2)}$ in Eq.\,(\ref{eqG2}) have the same 
functional form of the corresponding QCD
ones
\begin{align}
\label{g1qed}
g'^{(1)}(\alpha L) &= \f{A_q'^{(1)}}{\beta_0'} \f{\pla+\ln(1-\pla)}{\pla} \;\;,  \\
\label{g2qed}
g_N'^{(2)}(\alpha L) %\!\left(\alpha \tilde L;\f{M^2}{\mu^2_R},\f{M^2}{Q^2} \right)
&= %\f{{\overline B}_N^{(1)}}{\beta_0'}
\f{{\widetilde B}_{q,N}'^{(1)}}{\beta_0'} %\left(
   %+ A_q'^{(1)} \ln\f{M^2}{Q^2}
%\right) 
\ln(1-\pla) -\f{A_q'^{(2)}}{\beta_0'^2} 
\left( \f{\pla}{1-\pla} +\ln(1-\pla)\right) \nn \\ 
%&+ \f{A_q'^{(1)}}{\beta_0'} 
%\left( \f{\pla}{1-\pla} +\ln(1-\pla)\right) \ln\f{Q^2}{\mu_R^2}  \nn \\
& +\f{A_q'^{(1)} \beta_1'}{\beta_0'^3} \left( \f{1}{2} \ln^2(1-\pla)+ 
\f{\ln(1-\pla)}{1-\pla} + \f{\pla}{1-\pla}  \right) \;,  
\end{align}
while the novel function $g'^{(1,1)}(\as L,\alpha L)$,  which controls the mixed QCD-QED correction at leading logarithmic accuracy,
reads 
\begin{align}
\label{g11}
g'^{(1,1)}(\alpha_S L,\alpha L) &= \f{A_q^{(1)}\beta_{0,1}}{\beta_0^2\beta_0'} \,h(\lambda,\lambda')
+\f{A_q'^{(1)}\beta_{0,1}'}{\beta_0'^2\beta_0} \,h(\lambda',\lambda) \;\;,  \\
\end{align}
with 
\begin{align}
\label{hfun}
h(\lambda,\lambda')&=-\frac{\lambda'}{\lambda-\lambda'}\ln(1-\lambda)
+\ln(1-\lambda')\left[\frac{\lambda(1-\lambda')}{(1-\lambda)(\lambda-\lambda')}
+\ln\left(\frac{-\lambda'(1-\lambda)}{\lambda-\lambda'}\right)\right]\nn\\
&-\rm{Li_2}\left(\frac{\lambda}{\lambda-\lambda'}\right)+Li_2\left(\frac{\lambda(1-\lambda')}{\lambda-\lambda'}\right),
\end{align}
where $\lambda=\frac{1}{\pi}\beta_0\, \alpha_S \, L$, $\lambda'=\frac{1}{\pi}\beta_0'\, \alpha \, L$, and
$\beta_0$, $\beta_0'$, $\beta_1'$, $\beta_{0,1}$, $\beta_{0,1}'$ are the coefficients of the QCD and QED $\beta$ functions:
%the first two %(renormalization scheme independent) 
%coefficients of the pure QED $\beta$ function:
\begin{align}
\label{bqcd}
\frac{d\ln \alpha_S(\mu^2)}{d\ln\mu^2}=\beta(\alpha_S(\mu^2),\alpha(\mu^2))
=
-\sum_{n=0}^{\infty}\beta_n\left( \frac{\alpha_S}{\pi}\right)^{n+1}
-\sum_{n,m+1=0}^{\infty}\beta_{n,m}\left( \frac{\alpha_S}{\pi}\right)^{n+1}\left( \frac{\alpha}{\pi}\right)^{m}\,,
\end{align}
\begin{align}
\label{bqed}
\frac{d\ln \alpha(\mu^2)}{d\ln\mu^2}=\beta'(\alpha(\mu^2),\alpha_S(\mu^2))=
-\sum_{n=0}^{\infty}\beta_n'\left( \frac{\alpha}{\pi}\right)^{n+1}
-\sum_{n,m+1=0}^{\infty}\beta_{n,m}'\left( \frac{\alpha}{\pi}\right)^{n+1}\left( \frac{\alpha_S}{\pi}\right)^{m}\,.
\end{align}
In Eqs.\,(\ref{bqcd},\ref{bqed}) we have consistently included the mixed QCD-QED  contributions to the running of 
the QCD and QED couplings through the coefficient $\beta_{n,m}$ and $\beta_{n,m}'$.
The explicit expressions of the  coefficients $\beta_0$, $\beta_0'$, $\beta_1'$, $\beta_{0,1}$, $\beta_{0,1}'$ are:
\begin{align}
\beta_0 &= \frac{1}{12}(11\,C_A-2\,n_f)\, ,\qquad \beta_{0,1}=-\frac{N_q^{(2)}}{8}\,, \\
\beta_0' &= -\frac{N^{(2)}}{3}\, ,\qquad
\beta_1' = -\frac{N^{(4)}}{4}\,,\qquad \beta_{0,1}'=-\frac{C_FC_AN_q^{(2)}}{4},
\end{align}
where we have defined
\begin{align}
N^{(n)} &= N_c \sum_{q=1}^{n_f} e_q^n + \sum_{l=1}^{n_l} e_l^n\,,\\
N_q^{(n)} &= \sum_{q=1}^{n_f} e_q^n \,
\label{eqnf}
\end{align}
with  $N_c=3$ the number of colours, $C_A=N_c$, $C_F=(N_c^2-1)/(2N_c)$,  $n_f$ ($n_l$) the number of quark (lepton) flavours and $e_q$ ($e_l$) the quark (lepton) 
electric charges ($e_q=2/3$ for up-type quarks, $e_q=-1/3$ for down-type quarks, $e_l=1$ for leptons). 
%The sums 
%in Eq.\,(\ref{eqnf}) are performed over the number of 
%active quark and lepton flavours,
%with electric charge $e_q$ and $e_l$ respectively.
%

The coefficients $A_q'^{(1)}$  and $A_q'^{(2)}$,
which have been obtained  from the corresponding coefficients 
in QCD~\cite{Kodaira:1981nh,Catani:1988vd}
through an Abelianization 
algorithm~\cite{deFlorian:2015ujt,deFlorian:2016gvk},
read
\begin{align}
A_q'^{(1)} &= e_q^2 \, ,\qquad
A_q'^{(2)} = -\frac{5}{9}\,e_q^2 \,N^{(2)}
%\\ C^{(0,1),{\rm N-ind}}_q &=& \frac{e_q^2}{2}\left(\frac{\pi^2}{2}-4\right) \, .
\end{align}
while the coefficient $\widetilde B_{q,N}'^{(1)}$ is
\begin{align}
\widetilde B_{q,N}'^{(1)} = B_{q}'^{(1)} + 2\gamma_{qq,N}'^{(1)}\,,
\end{align}
with
\begin{align}
  B_q'^{(1)} &= -\frac{3}{2} \, e_q^2 \,,\\ 
\gamma_{qq,N}'^{(1)}&=e_q^2\,\left(\frac34+\frac{1}{2N(N+1)}-\gamma_E-\psi_0(N+1)\right)\,,\\
\gamma_{q\gamma,N}'^{(1)}&=\frac32\,e_q^2\,\frac{N^2+N+2}{N(N+1)(N+2)}\,.
\end{align}
where $\psi_0(N)$ is the digamma function and $\gamma_{ab,N}'^{(1)}$ are the leading-order (LO) anomalous dimensions in 
QED~\footnote{The QED anomalous dimension $\gamma_{q\gamma,N}'^{(1)}$ is required at the NLL in the general multiflavour case 
(see Appendix A of Ref.\,\cite{Bozzi:2005wk}).}.
%requires also the knowledge of the other anomalous dimensions in QED.
%$\gamma_{N}'$. 
%In particular at the NLL it is required the anomalous dimension $\gamma_{q\gamma,N}'^{(1)}$
%\begin{align}
%\gamma_{q\gamma,N}'^{(1)}=\frac32\,e_q^2\,\frac{N^2+N+2}{N(N+1)(N+2)}\,.
%\end{align}

The LL mixed QCD-QED corrections are included in the $g'^{(1,1)}(\as L,\alpha L)$  function which depends on
the coefficients $A_q^{(1)}=C_F$,  $A_q'^{(1)}$, $\beta_0$, $\beta_0'$, $\beta_{0,1}$, $\beta_{0,1}'$. We note that
these logarithmic terms do not contribute to the $\mathcal{O}(\alpha \alpha_S)$ fixed order corrections and their dominant contribution is
of the order $\alpha \alpha_S^2 L^3$.

In order to reach the complete NLL+NLO accuracy for QED, the functions ${\cal H}_{a_1a_2,N}'^{F \,(1)}$
and the finite component of the partonic cross sections $d \hat{\sigma}_{a_1a_2}^{({\rm fin.})}$ at the first non-trivial order are needed.
%for the various partonic channles $a_1a_2\to F$ are also needed.
%The Abelianization of corresponding QCD functions (see Eqs.\,(9)-(10) of Ref.\cite{Bozzi:2008bb})  gives
The hard-collinear functions ${\cal H}_{a_1a_2,N}'^{F \,(1)}$ read
\begin{align}
\label{h1qed}
{\cal H}_{q\bar q \leftarrow q\bar q,N}'^{F \,(1)}&=\frac{e_q^2}{2} \, \left(\frac{2}{N(N+1)} -8 + \pi^2  \right) \,,\\
{\cal H}_{q\bar q \leftarrow \gamma q,N}'^{F \,(1)}&={\cal H}_{q\bar q \leftarrow q\gamma ,N}'^{F \,(1)}=\frac{3\, e_q^2}{(N+1)(N+2)} \,,\\
{\cal H}_{q\bar q \leftarrow \gamma \gamma,N}'^{F \,(1)}&={\cal H}_{q\bar q \leftarrow q q ,N}'^{F \,(1)}={\cal H}_{q\bar q \leftarrow \bar q \bar q ,N}'^{F \,(1)}\,=0.
\end{align}

The explicit perturbative scale dependence of the formulae in Eqs.\,(\ref{g1qed},\ref{g2qed},\ref{h1qed}) is the same as the corresponding one
in QCD (see Eqs.\,(22,23,46) of Ref.\cite{Bozzi:2005wk}).

\section{Numerical results}
\label{sec:results}
In this Section we present selected numerical results, by explicitly considering the transverse-momentum distribution of on-shell $Z$ bosons in hadronic collisions.
We start our analysis from the resummed results in QCD at NNLL+NNLO as implemented in the \texttt{DYqT} numerical code
\cite{Bozzi:2008bb,Bozzi:2010xn} and we include the QED corrections up to NLL+NLO.

As for the electroweak couplings, we use the %the so-called $G_\mu$ scheme, {\bf (CHANGE with the 
following input parameters:
%$G_F = 1.16637\times 10^{-5}$~GeV$^{-2}$,
$\alpha(m_Z^2)=1/127.95$, $\sin^2\theta_W=0.23129$ and
$m_Z = 91.1876$~GeV~\cite{Patrignani:2016xqp}.  The electromagnetic coupling $\alpha$ is evaluated at 1-loop and 2-loops respectively at LO and NLO in QED. 
%within the $\rm{\overline{MS}}$ renormalization scheme.
%, $\Gamma_Z=2.4952$~GeV, $m_W = 80.399$~GeV 
%and $\Gamma_W=2.085$~GeV. {\bf (CURRENT VALUES (PDG 2016) GF = 1.1663787 and mW=80.385, CHECK IT)} 
The hadronic cross section is computed using the \texttt{NNPDF3.1QED} parton distribution function (PDF) set at NNLO in QCD~\cite{Ball:2017nwa,Bertone:2017bme},
which includes the photon PDF as determined within the \texttt{LUX} method \cite{Manohar:2016nzj,Manohar:2017eqh} as well as the LO QED effects in the
evolution of the parton densities. 
The strong coupling $\alpha_S$ is evaluated at 3-loop order with $\alpha_S(m_Z^2)=0.118$ within the $\rm{\overline{MS}}$ renormalization scheme.
We work with $n_l=3$ charged leptons and $n_f=5$ flavours of light quarks
in the massless approximation.
%and $N_f=5$ flavours of light quarks in the massless approximation, 
% with 
%$\alpha(m_Z^2)=1/132.358$. % determined, within the $G_\mu$ scheme, by $\alpha=\sqrt{2}\,G_F\,m_W^2\,(1-m_W^2/m_Z^2)/\pi$
%{\bf (%Is $\alpha$ defined in the on-shell or MSbar scheme? 
%Mention effects of hadronic vacuum polarisation contributions to $\alpha$?)}

We set the central value of the renormalization, factorization and resummation scales
at $\mu_R=\mu_F=2Q=m_Z$. We provide an estimate of the perturbative uncertainties of the calculation due to missing higher-order QED
terms performing the variation of the renormalization ($\mu_R'$) and resummation ($Q'$) scales associated to the QED contributions.
Specifically, the variation of $\mu_R'$ and $Q'$ around their central value can be used to estimate the effects of yet uncalculated 
QED contributions respectively  at fixed order and logarithmic accuracy level.
We thus perform an independent variation of $\mu_R'$ and $Q'$ in the range  $m_Z/2\leq\{\mu_R',2Q'\}\leq 2m_Z$ with the constraint $1/2\leq\{\mu_R'/Q'\}\leq 2$.
%For what concern the factorization scale, 
In principle it would be possible to consider an independent variation also for the factorization scale ($\mu_F'$) related to the QED emissions.
%separated by the QCD factorization scale $\mu_F$. 
However since the QCD and QED factorization scales are assumed to be equal in the evolution of parton densities~\cite{Bertone:2017bme}
we keep the factorization scale 
fixed at the central value $\mu_F'=\mu_F=m_Z$
in our numerical study. The dependence of the 
resummed QCD predictions from factorization, renormalization and resummation 
scales has been studied in detail in Refs.\cite{Bozzi:2008bb,Bozzi:2010xn,Catani:2015vma}.

%%====================================
\begin{figure}[t]
\begin{center}
\begin{tabular}{cc}
\includegraphics[width=0.46\textwidth]{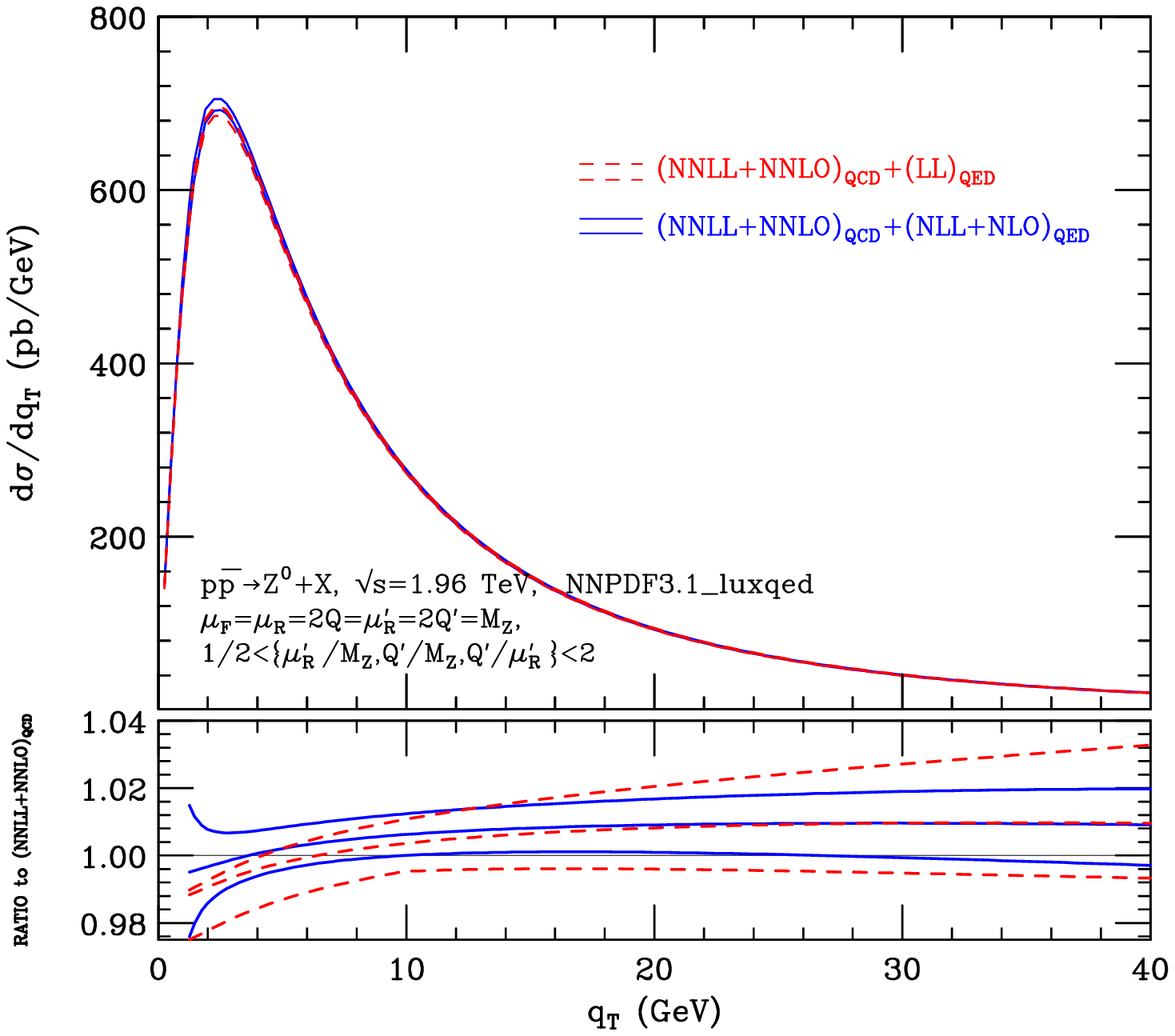} & \includegraphics[width=0.46\textwidth]{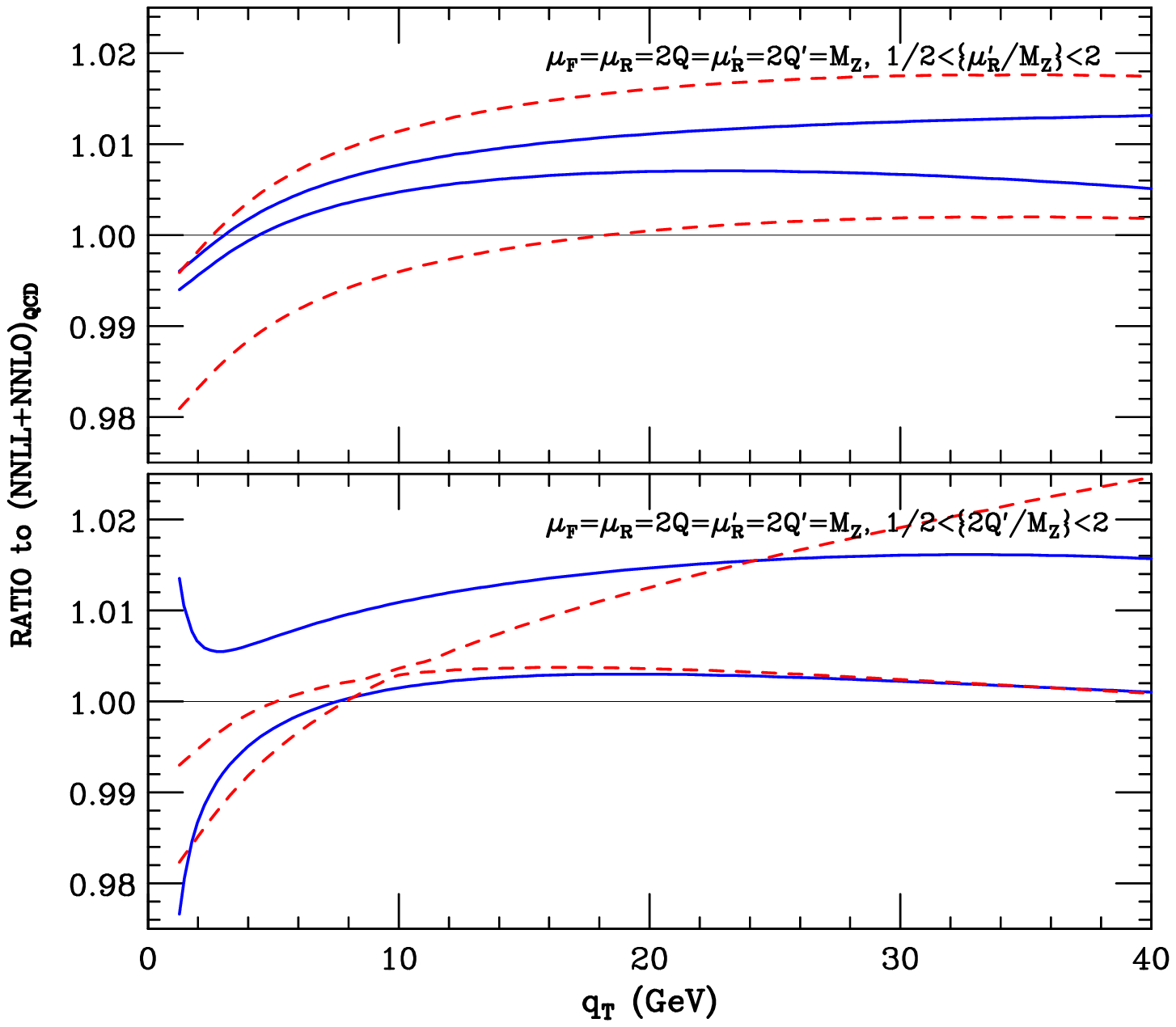}\\
\end{tabular}
\end{center}
%\vspace*{-.1cm}
\caption{\label{fig1}
{\em 
The $q_T$ spectrum of $Z$ boson at the Tevatron ($\sqrt{s}=1.96$~TeV). 
Left panel: NNLL+NNLO QCD results are combined with the LL (red dashed) 
and NLL+NLO (blue solid) QED effects together with the corresponding QED uncertainty bands. The bands are obtained by performing the variation of the $\mu_R'$
and $Q'$ scales  in QED around their central value as described in the text. 
The lower panel shows the ratio of the QED scale-dependent results with respect to the standard NNLL+NNLO QCD result at central value of the scales.
Right panel: upper (lower) panel shows the ratio of the resummation (renormalization) QED scale-dependent results with respect to the standard 
NNLL+NNLO QCD result at central value of the scales.
}}
\end{figure}
%%====================================
%\clearpage

We start considering on-shell $Z$ production in $p\bar p$ collisions at Tevatron Run II energies ($\sqrt{s}=1.96$~TeV). In Fig.\,\ref{fig1} (left panel)
we present the NNLL+NNLO QCD results combined with the LL (red dashed) and NLL+NLO (blue solid) QED corrections. The lower panel
presents the ratio of our predictions with respect to the standard NNLL+NNLO QCD result at the corresponding central scale. 

We observe that the resummation of the QED contributions at LL accuracy for central scale values ($\mu_R'=2Q'=m_Z$) has the effect to make 
the $q_T$ spectrum
slightly harder. The QED effects at LL are negative in the small $q_T$ region ($q_T\ltap 6$~GeV)  while they are positive in the $q_T$ region 
$6 \ltap q_T\ltap 40$~GeV. The impact of these effects reaches the $\mathcal{O}(1\%)$ level.  
We recall that, owing to the unitarity constraint of the
resummation formalism that we used~\cite{Bozzi:2005wk}, the LL QED effects for central values of the scales give vanishing contribution to the total 
(i.e.\ $q_T$ integrated) cross section: they only affect the
{\itshape shape} of the $q_T$ distribution by (slightly) {\itshape shifting} part of the cross section to higher values of $q_T$. 
This physical effect to the $q_T$ distribution is expected and it is due to the contribution to the $Z$ recoil generated 
by soft and collinear photon emissions to all orders in QED.
The NLL+NLO QED effects at central value of the scales $\mu_R'=2Q'=m_Z$ are of $\mathcal{O}(0.5\%)$ level and they are positive for the entire $q_T$ region 
considered ($q_T\ltap 40$~GeV). Thanks to the perturbative unitarity of the resummation formalism, the contribution to the integrated $q_T$ distribution  
of the NLL+NLO QED result exactly coincides 
with the NLO QED correction
to the total cross section which is equal to $+0.3\%$.

%%%%  FIG2

%%====================================
\begin{figure}[t]
\begin{center}
\begin{tabular}{cc}
\includegraphics[width=0.46\textwidth]{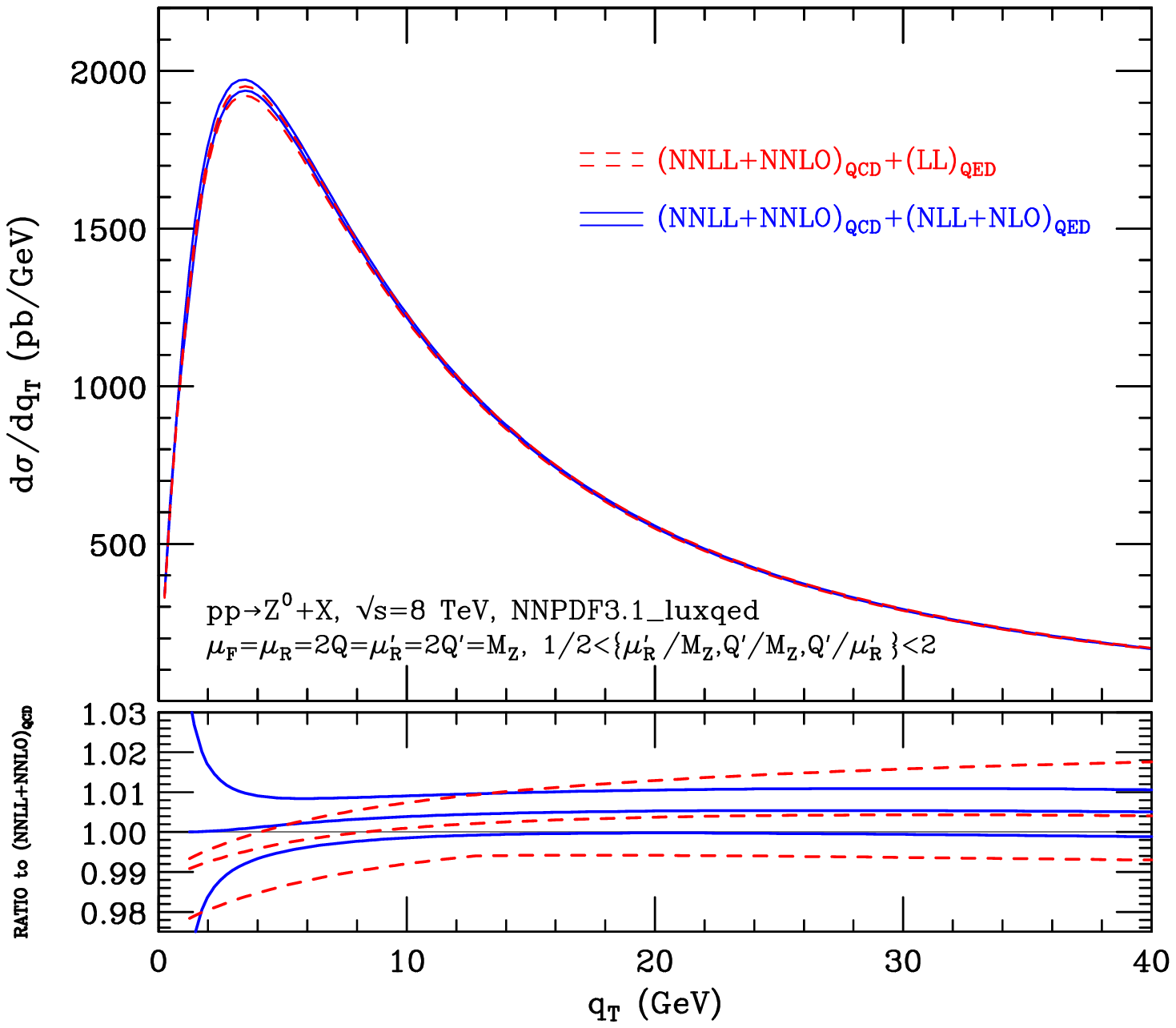} & \includegraphics[width=0.46\textwidth]{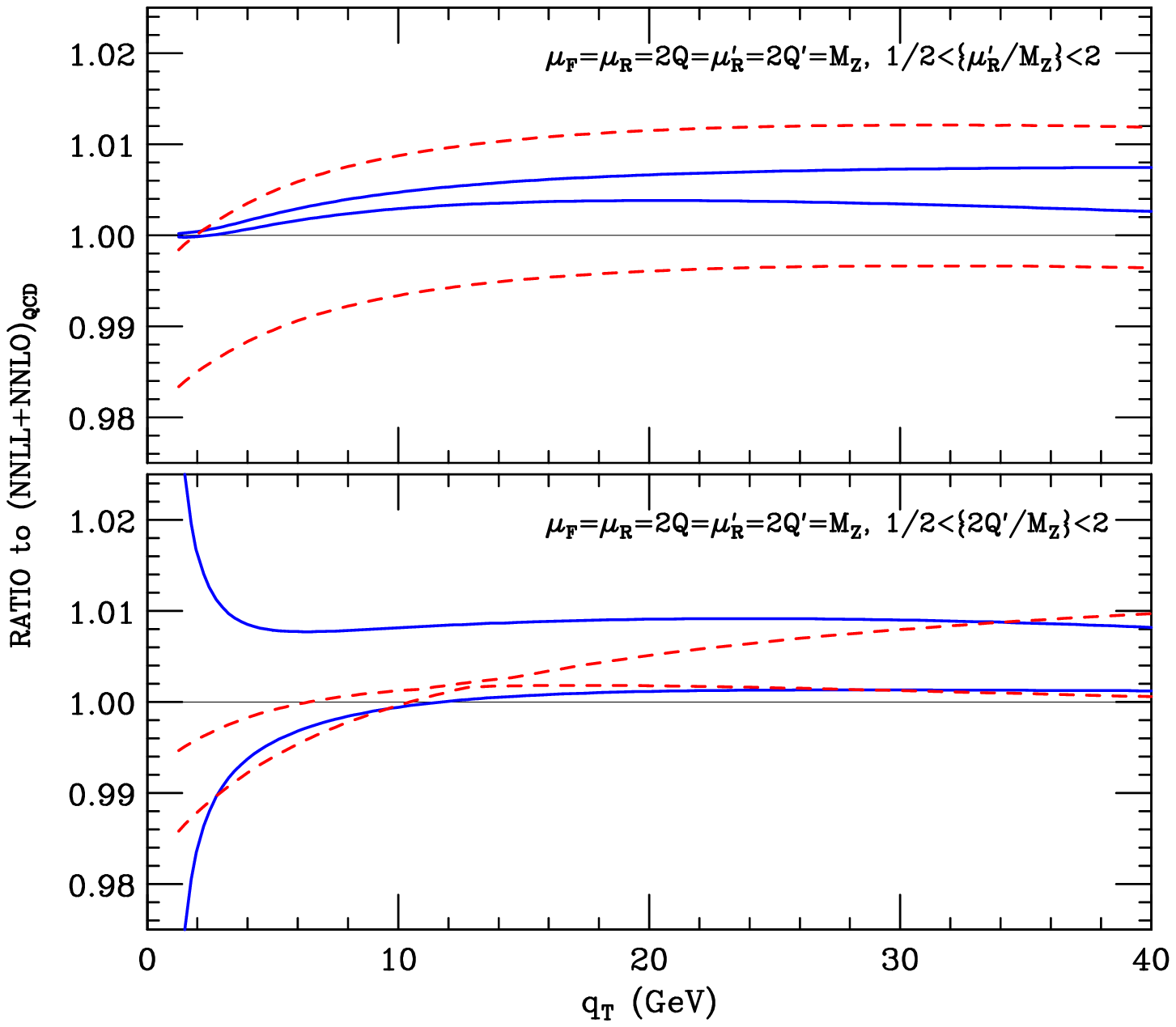}\\
\end{tabular}
\end{center}
%\vspace*{-.1cm}
\caption{\label{fig2}
{\em 
The $q_T$ spectrum of $Z$ boson at the LHC ($\sqrt{s}=8$~TeV). 
Left panel: NNLL+NNLO QCD results are combined with the LL (red dashed) 
and NLL+NLO (blue solid) QED effects together with the corresponding QED uncertainty bands. The bands are obtained 
as in Fig.~\ref{fig1}.
%by performing the variation of the $\mu_R'$
%and $Q'$ QED scales (as described in the text) around the central value $m_Z/2$. 
%The lower panel shows the ratio of the QED scale-dependent results with respect to the standard NNLL+NNLO QCD result at central value of the scales.
%Right panel: upper (lower) panel shows the ratio of the resummation (renormalization) QED scale-dependent results with respect to the standard 
%NNLL+NNLO QCD result at central value of the scales.
}}
\end{figure}
%====================================

%%%%

By considering the scale variation band, we observe that the LL QED effects 
have an uncertainty of around $2\%$ in the small $q_T$ region
($q_T\ltap 6$~GeV)
which increases up to $4\%$ in the intermediate $q_T$ region ($30 \ltap q_T\ltap 40$~GeV). 
The inclusion of the NLL+NLO QED corrections reduces the scale variation band 
by roughly a factor 2.
In order to better illustrate the dependence on the QED scales, in Fig.\,\ref{fig1} (right panel) we separately consider $\mu_R'$ and $Q'$ variations. 
The upper plot corresponds to the variation of $\mu_R'$ by a factor 2 around the central value and the lower plot corresponds to an analogous
variation of $Q'$. As expected from the QED running of $\alpha$, the $q_T$ distribution decreases (increases) by decreasing (increasing) the value of
$\mu_R'$. The $\mu_R'$ scale dependence is flat and at the level of $2\%$ at LL and it decreases to around $0.5-1\%$ at NLL+NLO. 
%In particular, we observe
%that the reduction of the $\mu_R'$ dependence is particularly significative in the small $q_T$ region, where our calculation is an effective NLO QED calculation. 
%Owing to the unitarity constraint of the resummation formalim, 
%The $Q'$ dependence does not modify the total (i.e.\ $q_T$ integrated) cross section  and it
%affects only the {\itshape shape} of the $q_T$ distrubution. 
%Owing to the unitarity constraint of the resummation formalim, 
%Thanks to the unitarity constraint, the $Q'$ dependence does not modify the total cross section  and it
%affects only the {\itshape shape} of the $q_T$ distrubution. 
%As discussed above the $Q'$ dependence does not affect the integral of the $q_T$ distribution. 
The $Q'$ dependence does not affect the integral of the $q_T$ distribution which is constrained by perturbative unitarity. 
The resummation scale dependence at LL is about $1\%$ 
for $q_T\ltap 5$~GeV, it reduces up to $0.1\%$ for $q_T\sim 10$~GeV and it increases again up to about $2\%$ for $q_T\sim 30-40$~GeV. The reduction
of the $Q'$ dependence in the region $q_T\sim 10$~GeV is due to the crossing (necessary in order to fulfill unitarity) between the predictions with $Q'=m_Z$ and $Q'=m_Z/4$. This accidental
cancellation suggests that the $Q'$ variation band at LL might
underestimates the perturbative uncertainty of the prediction in the region around $q_T\sim 10$~GeV.
In fact at NLL+NLO  the $Q'$ dependence for $q_T\gtap 4$~GeV is constant and 
at  $\mathcal{O}(1\%)$ level. 
Below the peak the $Q'$ dependence rapidly increases. We note that 
in the region $q_T\ltap 3$~GeV
the $q_T$ distribution is steeply falling to zero.
We finally observe that at LL the scale variation is mainly driven
by $\mu_R'$ because the Born-level cross section is proportional to $\alpha$,
while at NLL+NLO the scale uncertainty is dominated by the $Q'$ dependence.
%, due to Sudakov suppression of the real radiation,
%the non perturbative QCD effects start to be dominant. 
%therefore 
%and we do not expect the scale dependence to be a trustable estimate of the theoretical uncertainty. 
%{\bf (NOT SURE ABOUT THE ORIGIN OF LARGE Q' DEPENDENCE AT VERY SMALL QT)} 
%Moreover  in this region the non perturbative QCD effects cannot be regarded as small corrections and they have to be porperly taken into account. 
%therefore 

%The increasing of the scale variation band at 

The integral over $q_T$ of the predictions with NLL+NLO QED effects are in agreement 
(for fixed values of renormalization and factorization scales) with the corresponding
NLO QED contributions to the total cross section at the $\mathcal{O}(0.1\%)$ level~\footnote{The NLO QED ($\mathcal{O}(\alpha^2)$) corrections 
to the total cross section and to the finite component $d \hat{\sigma}^{({\rm fin.})}$ of the $q_T$ distribution
have been obtained through the Abelianization of the calculations in 
Ref.\cite{Hamberg:1990np} and in Ref.\cite{Gonsalves:1989ar} respectively.}, 
thus checking the numerical accuracy of our implementation.

%\clearpage

%====================================
\begin{figure}[t]
\begin{center}
\begin{tabular}{cc}
\includegraphics[width=0.46\textwidth]{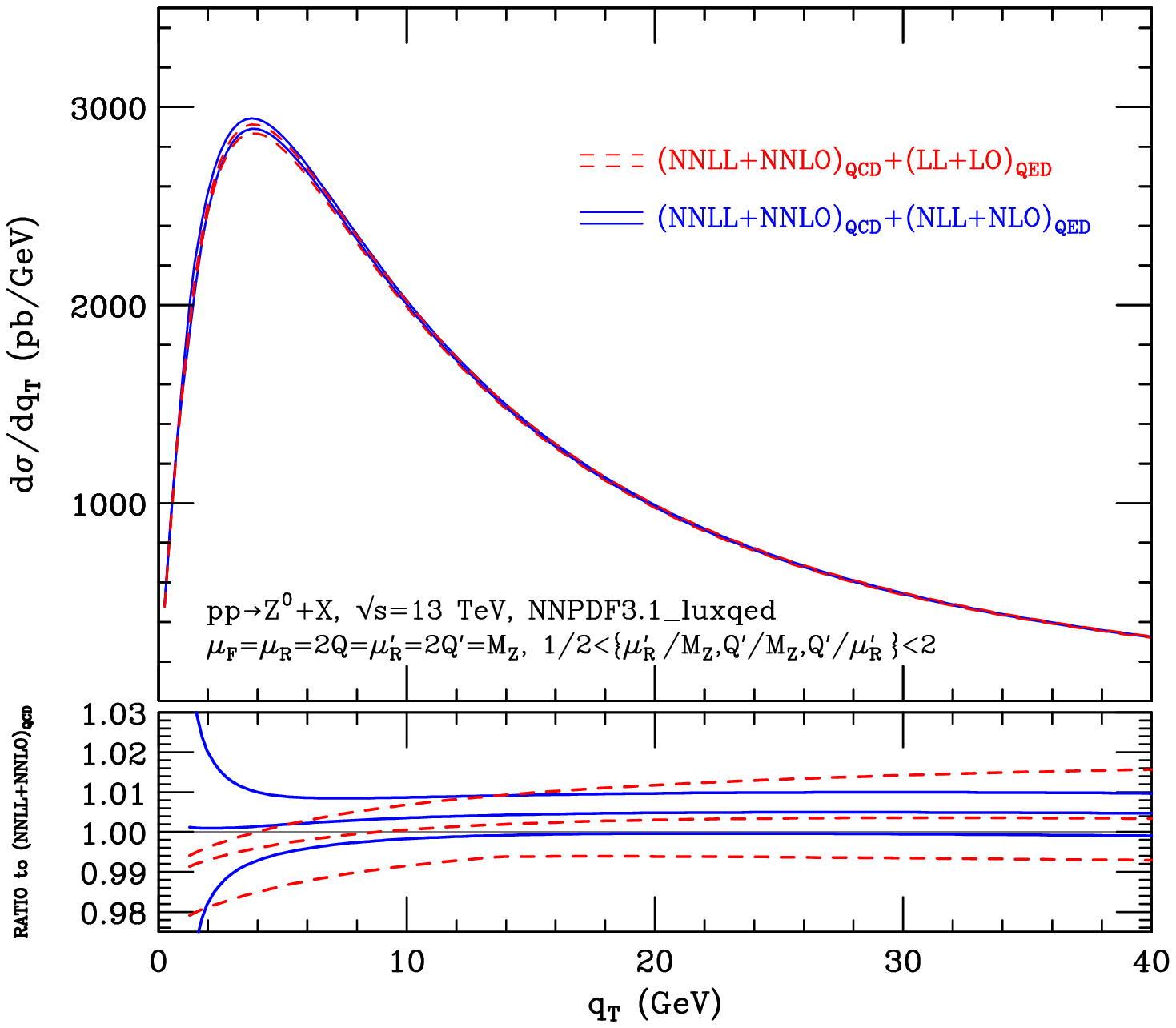} & \includegraphics[width=0.46\textwidth]{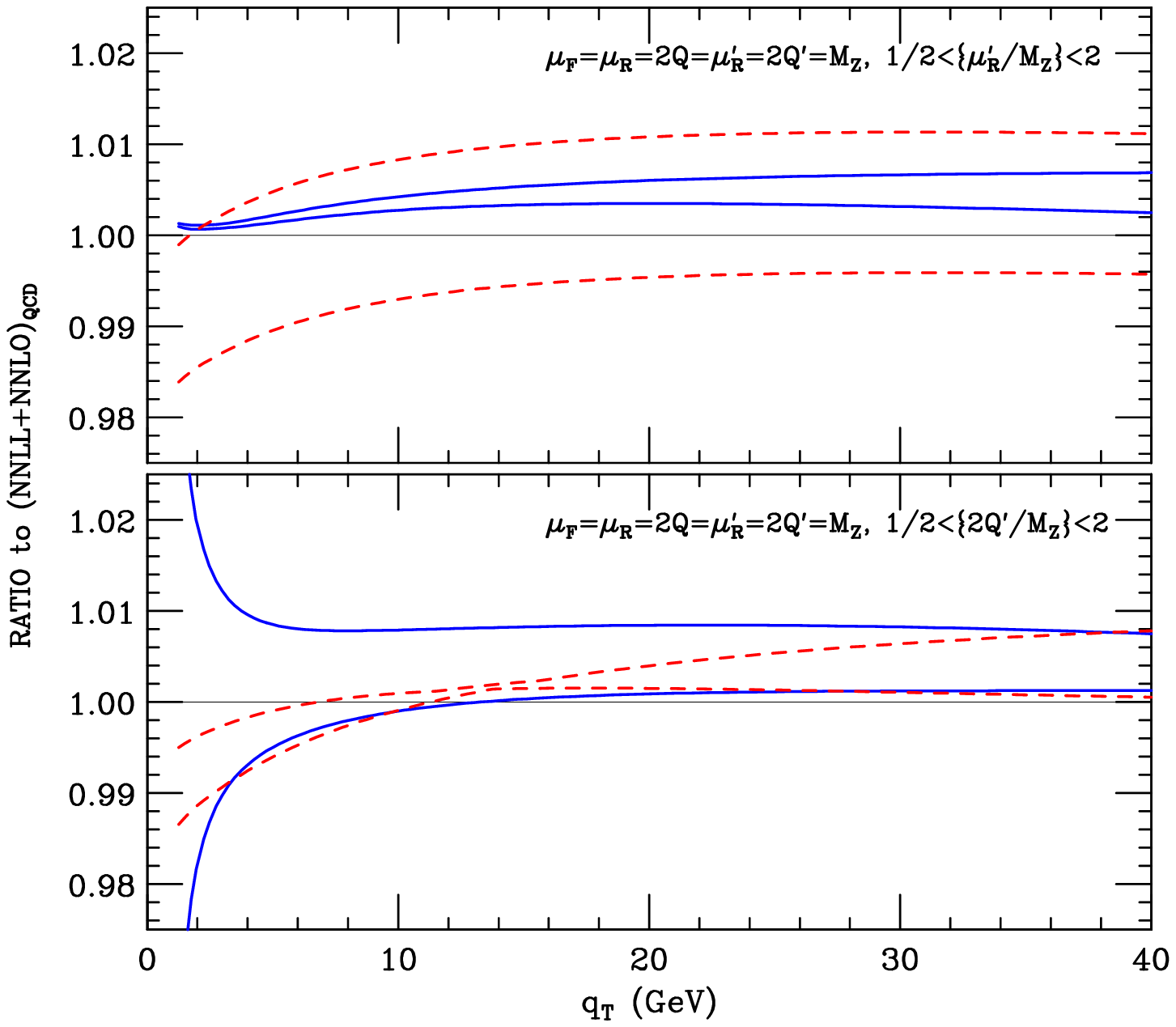}\\
\end{tabular}
\end{center}
%\vspace*{-.1cm}
\caption{\label{fig3}
{\em 
The $q_T$ spectrum of $Z$ boson at the LHC ($\sqrt{s}=13$~TeV). 
Left panel: NNLL+NNLO QCD results are combined with the LL (red dashed) 
and NLL+NLO (blue solid) QED effects together with the corresponding QED uncertainty bands. The bands are obtained 
as in Figs.~\ref{fig1} and \ref{fig2}.
%by performing the variation of the $\mu_R'$
%and $Q'$ QED scales (as described in the text) around the central value $m_Z/2$. 
%The lower panel shows the ratio of the QED scale-dependent results with respect to the standard NNLL+NNLO QCD result at central value of the scales.
%Right panel: upper (lower) panel shows the ratio of the resummation (renormalization) QED scale-dependent results with respect to the standard 
%NNLL+NNLO QCD result at central value of the scales.
}}
\end{figure}
%%====================================

We now turn to consider on-shell $Z$ production in $pp$ collisions at the LHC energies. % ($\sqrt{s}=7$~TeV and $\sqrt{s}=14$~TeV, respectively). 
 In Fig.\,\ref{fig2} (left panel) we present our results for LHC Run I ($\sqrt{s}=8$~TeV) and in
Fig.\,\ref{fig3} (left panel) the results for LHC Run II ($\sqrt{s}=13$~TeV).
The effects of the QED contributions at the LHC 
are qualitatively similar and quantitatively slightly smaller with respect to the case of the Tevatron.
At the LHC a lower sensitivity to QED contributions is expected due to the enhanced importance of the 
$qg$ channel with respect to the $q\bar q$ channel.

% very similar between
%the case of LHC Run I and Run II  energies.
%This fact is expected since 
From the left panels of Figs.\,\ref{fig2} and \ref{fig3}
we observe that the QED effects at LL accuracy, 
for central value of the scales,
give a negative $\mathcal{O}(1\%)$ contribution in the small $q_T$ region 
($q_T\ltap 8-10$~GeV)  and a positive $\mathcal{O}(0.5\%)$ contribution 
in the $q_T$ region 
$10 \ltap q_T\ltap 40$~GeV. 
The NLL+NLO QED effects are at the level of $\mathcal{O}(0.5\%)$
and positive for the entire $q_T$ region 
considered ($q_T\ltap 40$~GeV). 
The NLO QED contribution to the 
 total cross section is equal to $+0.3\%$.

By considering the perturbative uncertainty
shown in the lower-left panels of Figs.\,\ref{fig2} and \ref{fig3},
we observe that the LL QED effects have an uncertainty 
of around $2\%$ in the small $q_T$ region
($q_T\ltap 10$~GeV)
which increases up to $2.5-3\%$ in the intermediate $q_T$ region 
($30 \ltap q_T\ltap 40$~GeV). 
The inclusion of the NLL+NLO QED corrections reduces the scale variation band 
by roughly a factor 1.5-2.
As in the Tevatron case the QED uncertainty is dominated by the 
renormalization (resummation) scale dependence at LL (NLL+NLO).

%\clearpage
\section{Summary}
\label{sec:conclusions}

In this paper we 
%have considered the transverse-momentum ($q_T$) distribution of on-shell $Z$ bosons produced in hadronic collisions.
%We 
have extended the QCD transverse-momentum ($q_T$) resummation formalism of Refs.~\cite{Catani:2000vq,Bozzi:2005wk} 
in order to deal with the simultaneous emission of QCD and QED initial state  parton radiation. 
In particular, through the application of the Abelianization algorithm of Refs.~\cite{deFlorian:2015ujt,deFlorian:2016gvk},
we have obtained  analytic resummed  results which are valid for the production of generic neutral and colourless high-mass systems
in hadronic collisions.

We have applied our formalism to the
the transverse-momentum ($q_T$) distribution of on-shell $Z$ bosons produced at the Tevatron and the LHC.
Starting from the known QCD results at
next-to-next-to-leading logarithmic accuracy (NNLL) 
and next-to-next-to-leading order (NNLO) (i.e.\ $\mathcal{O}(\alpha_S^2)$)~\cite{Bozzi:2008bb,Bozzi:2010xn},
we have included the resummation
of the logarithmically enhanced QED contributions at small values of $q_T$
up to next-to-leading logarithmic  accuracy (NLL)
and of  the mixed QCD-QED contributions at leading logarithmic accuracy  (LL). 
Our results 
have been consistently matched with the next-to-leading
fixed-order (NLO) results (i.e.\ $\mathcal{O}(\alpha^2)$)
in QED at small, intermediate and large values of $q_T$.

We have found that the resummation of the QED contributions at leading-logarithmic accuracy
has the effect of slightly shifting (roughly $\mathcal{O}(\pm 1\%)$) 
part of the cross section to higher values of $q_T$. 
The NLL+NLO QED effects at the central value of the scales are 
$\mathcal{O}(+1\%)$ and rather flat for the entire $q_T$ region 
considered ($q_T\ltap 40$~GeV). 

We have then performed a study of the scale dependence of the QED contribution in order to estimate the corresponding perturbative uncertainty.
We have shown that the scale variation band at LL is 
of $\mathcal{O}(2-4\%)$ and that 
the inclusion of the NLL+NLO QED results reduces the scale variation band by roughly a factor 1.5-2.
The QED uncertainty is dominated by the renormalization and resummation scale dependence at LL and NLL+NLO respectively. 

In conclusion we have found that, given the present theoretical and experimental accuracy, the QED resummation effects for on-shell $Z$ boson production
are not negligible. In particular the inclusion of the NLL+NLO QED corrections reduces the effect of the renormalization scale ambiguity of
the QED coupling $\alpha$  by consistently including running coupling  effects.

The inclusion of the QED effects from the the leptonic decay of the $Z$ boson and the %(non trivial) 
generalization of our results to the case of
charged final states 
(e.g.\ $W^\pm$) are left for future investigations.

%\noindent {\bf Acknowledgements.}
\paragraph{Acknowledgements.} 

We gratefully acknowledge Stefano Catani, Daniel de Florian and Alessandro Vicini for useful discussions and comments on the manuscript.
G.\,S.\ would like to thank also Germ\'an Rodrigo and F\'elix Driencourt-Mangin for interesting discussions.
This research was supported in part by Fondazione Cariplo under the grant number 2015-0761
and by the COST Action CA16201 PARTICLEFACE.

%%%%%%%%%%%%
%\clearpage

\end{document}